\documentclass[12pt,a4paper]{article}

\usepackage{latexsym}
\usepackage{amsfonts}
\usepackage{amssymb}
\usepackage{amsmath}
\usepackage{theorem}
\usepackage{enumerate}
\usepackage{cite}
\usepackage{url}
\usepackage[dvips]{color}
\usepackage{graphicx}

\theoremstyle{plain}
\theorembodyfont{\itshape}
\newtheorem{thm}{Theorem}

\theorembodyfont{\upshape}

\newcommand{\proof}{\noindent {\bf Proof:} \hspace{0.1in}}
\newcommand{\qed}{\hfill\mbox{\raggedright $\Box$}\medskip}
\newcommand{\mydate}{
 \ifcase\month \or
 January\or February\or March\or April\or May\or June\or
 July\or August\or September\or October\or November\or December\fi
 \space \number\year}

\begin{document}

\title{ On the local form of static plane symmetric space-times in the presence of matter}
\author{L. G. Gomes}
\date{Instituto de Matem\'atica e Computa\c{c}\~ao \\
      Universidade Federal de Itajub\'a , Brazil}
\maketitle

\thispagestyle{empty}

\begin{abstract}
\noindent
For any configuration of a static plane-symmetric distribution of matter along space-time,
there are coordinates where the metric can be put explicitly as a functional of
the energy density and pressures. It satisfies Einstein equations as far as 
we require the  conservation of the energy-momentum tensor, which is the single ODE for 
self-gravitating hydrostatic equilibrium. As a direct application, 
a general solution is given when the pressures are linearly related to the energy density,
recovering, as special cases, most of known solutions of static plane-symmetric Einstein equations.
\end{abstract}

\section{Introduction}

 The gravitational field in the vicinity of a large and isolated distribution of matter is 
approximately static if the interval of time considered is small compared to the
characte-ristic time the system changes. As the matter distribution does not suffer any 
substantial change except in only one direction, it is also plane symmetric. 
This is the case, for example, when we are dealing with Newtonian gravitation 
close to the surface of a planet. In this case, define 
the height from the ground as $z$, the gravitational potential
as $\phi$ and the atmospheric mass distribution and pressure as $\rho$
and $p$, respectively. They are approximately functions of $z$ only. Besides 
an appropriate equation of state relating the later thermodynamic variables,
there is the hydrostatic equilibrium \cite{Landau6}
\begin{equation}\label{eq:ClassicalHydrostaticEquilibrium}
 \frac{dp}{dz}= - \rho \, \frac{d\phi}{dz}  \qquad ,
\end{equation}
with the Newtonian potential a solution of Poisson equation 
$\frac{d^2 \phi}{d z^2} = \frac{1}{2} \, \rho $. Here units are such that $ 8 \pi G = 1$ 
and $c=1$. 

The gravitational equation is simple enough to be integrated and 
give $\phi$ as
\begin{equation}\label{eq:ClassicalPotential}
 \phi(z) = g \, z  + \frac{1}{2} \int_0^z \, du \, \int_0^u \, dw \, \rho (w)   \qquad , 
\end{equation}
where $g$ is the constant acceleration of gravity in the absence of atmosphere.
Using eq.(\ref{eq:ClassicalHydrostaticEquilibrium}) and eq.(\ref{eq:ClassicalPotential})
we deduce the self-gravitating equation for hydrostatic equilibrium:
\begin{equation}\label{eq:ClassicalHydrostaticEquilibrium2}
 \frac{d}{dz} \left( \, \frac{1}{\rho} \, \frac{dp}{dz} \, \right) = - \frac{1}{2} \, \rho  \qquad .
\end{equation}

To set the stage for the relativistic generalization of the situation above, we start
with a four dimensional Lorentzian manifold $M$ admitting an isometric action of the 
transformation group of the Euclidean plane with $2$-dimensional space-like orbits.
Let  $e_1$ and $e_2$ stand for the induced killing vector fields corresponding to
translations in orthogonal directions. We also assume there is another killing vector 
field $e_0$ which is time-like, orthogonal to and commuting with $e_1$ and $e_2$. 
Choosing a curve $\gamma (z)$ such that $\dot{\gamma}(z) \neq 0$  is orthogonal
to $e_i$, $i=0,1,2$, and spreading it all around using the fluxes of the three 
killing vector fields, we define a coordinate system where $e_0 = \frac{\partial}{\partial t}$ ,
$e_1 = \frac{\partial}{\partial x}$ , $e_2 = \frac{\partial}{\partial y}$ and the metric 
is 
\begin{equation}\label{eq:PlaneMetric}
 ds^2 = g_{00}(z) \, dt^2 + g_{11}(z) \, (dx^2 + dy^2) + g_{33}(z) \, dz^2 \quad.
\end{equation}
We also assume that the energy-momentum tensor is equally symmetric, such that in these
coordinates it is diagonal 
\begin{equation}\label{eq:GeneralEMT}
(\, T^{\mu}_{\nu} \, ) = \text{\bf diag} \, \{\, \rho (z), - p_{||}(z), - p_{||}(z), -p(z) \, \}  
\quad . 
\end{equation}
Note that the energy density
$\rho = T_0^0$, the pressure parallel to the plane of symmetry
$p_{||}=- T_1^1 =- T_2^2$ and the pressure orthogonal to it $p =- T_3^3$  behave as functions
under a coordinate change of the type $u=u(z)$. This implies they are all defined globally 
in such space-time. To simplify our analysis, we set
the pressure difference $\delta p$ as
\begin{equation}\label{eq:PressureDif}
 \delta p = p - p_{||}  \quad .
 \end{equation}
Such scheme defines what we mean by a static plane symmetric space-time 
(see also \cite{SKMHH}, ch. 15). 

Unlike Newtonian physics, where the gravitational part is played by one potential and 
one linear Poisson equation, 
in the relativistic analogue we deal with three unknown functions playing the hole
of the potential, $g_{00}$, $g_{11}$ and $g_{33}$, and three non-linear independent Einstein 
equations, which are far from being easily integrated (see \cite{AmGr}, for instance). 
Indeed, the component $g_{33}$ is spurious as we can make it equal unit in
a simple coordinate change $du = g_{33}(z) \, dz$.

 Although we may identify matter as a fluid, in the relativistic approach it is natural
to consider a more complex and general structure of the energy-momentum tensor. Therefore
there is no need to consider an isotropic stress, that is, $\delta p = 0$. As an example,
for an electric charged plane \cite{AmGr} we have $\delta p = 2 p = -2 \rho$. 
As a result, our complementary equations of state are two in number.%
\footnote{By an equation of state we mean any one coming from the matter source,
which could be in the form of an algebraic equation, as it is usual in thermodynamics,
as a differential equation coming from a lagragian, etc.}
This makes our system well defined: we have five independent unknown functions,
$g_{00}$, $g_{11}$, $\rho$, $p$ and $\delta p$, related by five independent
equations: three of them coming from the gravitational part and the rest
from the sort of matter present in the system. In this paper, our main result is to give $g_{00}$, $g_{11}$ and $g_{33}$ in terms of
$\rho$, $p$ and $\delta p$ such that they satisfy Einstein equations, as we explain bellow.

Define the metric 
\begin{equation}\label{eq:GeneralMetric}
 ds^2 = e^{2 \phi} \, dt^2
 - e^{4 (z_2-z_1)\, \int_0^z  dw \, \frac{\phi'}{w -z_2}} \, \left(dx^2 + dy^2 \right)
 -  \frac{ 4 \,(z_2-z_1)\, (z-z_1) }{(z-z_2)^2 }\, \frac{(\phi ')^2}{p} \, dz^2 \quad ,
\end{equation}
where $\phi = \ln (g_{00})^\frac{1}{2}$ is the generalized gravitational potential
given by 
\begin{equation}\label{eq:DefinitionPotential}
  \phi =  \int_0^z \, dw  \, \frac{w - z_2}{w -z_1} \, \left( 
 \frac{p}{ (w-z_1) (\rho- p) + 
 (z_2-z_1) ( \rho + 7 p - 4 \delta p) } \right) \quad .
\end{equation}
We also assume that the derivative $\phi'$ is continuous and different from zero
in a neighbourhood of $z=0$. Thus $p \ne 0$ close to $z=0$. The arbitrary constants
$z_1$ and $z_2$ satisfy $z_2 > z_1$ and $z_1 \, p_0 < 0$, as $p_0$ is the value of $p$
at $z=0$. 

The conservation of the energy-momentum tensor (\ref{eq:GeneralEMT}) in the space-time 
with the metric above, 
$\nabla_\mu \, T^{\mu}_\nu  = 0$,  
turns out to be the generalization of the hydrostatic equilibrium relation
\begin{equation}\label{eq:SelfGravityConservationPhi}
\frac{dp}{dz}= - \left( \, \rho + p  + \frac{4 \,(z_2-z_1)}{z-z_2} \,\delta p \,\right) \, \frac{d\phi}{dz} \quad .
\end{equation}
This is commonly find in the literature for an unknown $\phi$ and $\delta p =0$ 
(\cite{Wein}, sec. 5.4). If we apply the definition of $\phi$ (eq.(\ref{eq:DefinitionPotential})),
we arrive at the relativistic self-gravitating hydrostatic equilibrium equation
\begin{equation}\label{eq:SelfGravityConservation}
\frac{dp}{dz}= - \left( \, 
\frac{ (z-z_2) \,(\rho + p)  + 4 \,(z_2-z_1)\,\delta p}{(z-z_1) \,(\rho - p)  +  \,(z_2-z_1)\,(\rho +7 p - 4 \delta p)}  \,\right)\frac{p}{z - z_1} \quad .
\end{equation}

 In this paper we show that the metric in eq.(\ref{eq:GeneralMetric}) and the energy-momentum 
tensor in eq.(\ref{eq:GeneralEMT}) satisfy Einstein equations , that is, 
$R^{\mu}_{\nu} -  \frac{1}{2} R \, \delta^{\mu}_{\nu} = T^{\mu}_{\nu}$, provided the conservation above is attained. Furthermore, the case when $\phi$ cannot be defined
as in eq.(\ref{eq:DefinitionPotential}) is also explained. These are presented in 
section \ref{sec:LocalForm}. In the following, we show how it simplifies the analysis of
space-times in the presence of fluids with the prescribed equations of state 
$p = (\gamma(\rho)-1)\, \rho$ and $\delta p = \epsilon(\rho)\, \rho$, giving in a 
straightforward and unified way most of the solutions studied so far, which fall 
in the category of $\gamma$ and $\epsilon$ constants. In the last section we make our final
remarks.

%
%
%
\section{The local form of the metric}\label{sec:LocalForm}
Our main result is given in the  following theorem:
%
\begin{thm}\label{thm:GeneralSolution}
The metric given in eq.(\ref{eq:GeneralMetric}) and the energy-momentum tensor 
in eq.(\ref{eq:GeneralEMT}) sa-tisfy Einstein equations in a neighbourhood of $z=0$,
provided the later attains 
the self-gravitating hydrostatic equilibrium given in eq.(\ref{eq:SelfGravityConservation}).
\end{thm}
%
%
\proof
We proceed straightforwardly in computing the Einstein tensor for the metric 
(\ref{eq:GeneralMetric}). Thus, the non-vanishing components of the
Levi-Civita connection are
\begin{equation*}
 \Gamma^t_{tz}= \phi'   \quad ; \quad  \Gamma^x_{xz}= \Gamma^y_{yz}= 2\,\frac{(z_2-z_1)\,\phi'}{z -z_2} 
 \quad ; \quad   \Gamma^z_{tt} = \frac{e^{2 \phi}\,(z-z_2)^2\,p}{4\,(z-z_1)\,(z_2-z_1) \, \phi'}
\end{equation*}
\begin{equation*}
\Gamma^z_{xx} = \Gamma^z_{yy} = - \frac{(z-z_2)\, p}{2  \, (z-z_1)\, \phi'}\,
\, e^{4 (z_2-z_1)\, \int_0^z  dw \, \frac{\phi'}{w -z_2}}
\quad ; \quad
\Gamma^z_{zz} =  \frac{1}{2}\, \frac{d}{dz}\,
\ln \left(\frac{ 4 \,(z_2-z_1)\, (z-z_1) }{(z-z_2)^2 }\, \frac{(\phi ')^2}{p} \, \right)
\end{equation*}
Through the rest of our proof we assume $\nabla_\mu \, T^{\mu}_\nu =0$, which is equivalent to
eq.(\ref{eq:SelfGravityConservationPhi}). Therefore, during the calculation of the curvature
$R^{\lambda \mu}_{\kappa \nu} = g^{\mu \mu'}\, 
(\partial_\kappa \, \Gamma_{\nu \mu'}^\lambda - \partial_\nu \, \Gamma_{\kappa \mu'}^\lambda +\ldots)$ we apply the identities 
\begin{equation}\label{eq:Identity1}
\frac{p'}{\phi'}= - \left( \, \rho + p  + \frac{4 \,(z_2-z_1)}{z-z_2} \,\delta p \,\right) 
\end{equation}
and 
\begin{equation}\label{eq:Identity2}
\frac{p}{\phi'} = \frac{z - z_1}{z -z_2} \, \left( (z-z_1) (\rho- p) + 
 (z_2-z_1) ( \rho + 7 p - 4 \delta p) \right) \quad ,
\end{equation}
the last one coming from the very definition of $\phi$. Then we obtain for its independent components
\begin{equation}
 R^{tx}_{tx} = R^{ty}_{ty} = \frac{(z-z_2) \, p}{2 (z-z_1)}  
 \quad ; \quad
 R^{tz}_{tz} = \frac{1}{2} \left(\rho + p - 2 \, \delta p \right) + \frac{z_2-z_1}{z-z_1}\,p 
\end{equation}
\begin{equation}
 R^{xy}_{xy} = \frac{z_2-z_1}{z-z_1}\,p 
 \quad ; \quad
 R^{xz}_{xz} = R^{yz}_{yz} = - \frac{1}{2}  \left( \rho + \frac{z_2-z_1}{z-z_1}\,p \right) \quad.
\end{equation}
For the Ricci tensor $R^{\mu}_{\nu} = R^{\lambda \mu}_{\lambda \nu}$ and scalar curvature we have
\begin{eqnarray*}
R^t_t =  \frac{1}{2} \left(\rho + 3\, p - 2 \, \delta p \right) \quad 
 \quad ; \quad
R^x_x = R^y_y =  - \frac{1}{2} \left(\rho - p \right) \quad
\\
R^z_z =  - \frac{1}{2} \left(\rho - p + 2\, \delta p \right) \, 
 \quad ; \quad
 R  =  - \rho + 3 \, p - 2 \, \delta p  \quad . 
\end{eqnarray*}
Hence Einstein equations hold for $ds^2$, as can be readily verified.\\[3mm]
\qed
%
%
 
It is not true that $ds^2$ in eq.(\ref{eq:GeneralMetric}) can be well defined 
for any kind of matter. In order to be complete, the following theorem explains
when such a scheme is not possible. 

\begin{thm}\label{thm:SpecialSolutions}
 If $\phi'$ is nowhere well defined or if it vanishes everywhere, then
around each point of this static plane symmetric space-time one of the following
holds, with $ds^2$ satisfying Einstein equations:
\begin{enumerate}
\item[(i)] \underline{$p = \rho = 0$\, , \,$\delta p = \delta p (z)$}. 
Defining
$\varphi = - \int_0^z dw \, \frac{w-z_1}{\delta p \, + (w-z_1)^2}$\quad ,
we have 
\begin{equation}\label{eq:SpecialSolution1}
 ds^2 = e^{2\, \varphi} \, dt^2
 - \, \left(dx^2 + dy^2 \right) 
 - \, \left( \frac{\varphi' \, dz}{z-z_1}\right)^2  \quad .
\end{equation}
\item[(ii)] \underline{$p = 0$\, , \,$\rho = 4\, \delta p$\, , \,$\delta p = \delta p (z)$}. 
Defining
$\varphi = \frac{1}{3} \, \int_0^z dw \, 
\frac{w-z_1}{3 \, \delta p\,+ (w-z_1)^2}$\quad ,
we have
\begin{equation}\label{eq:SpecialSolution2}
 ds^2 = e^{2 \,\varphi} \, dt^2
 - \, e^{-\,4\, \varphi} \, \left(dx^2 + dy^2 \right) 
 - \, \left( \frac{3\,\varphi' \, dz}{z-z_1}\right)^2  \quad .
\end{equation}
\item[(iii)] \underline{$\rho  = p\, , \,\delta p = 2\, p$}. There are constants 
$\alpha$ and $\beta$ such that
\begin{equation}\label{eq:SpecialSolution3.1}
 p =  \frac{4 \, \alpha \, \beta }{3 \,(1+(\, \alpha + \beta \,) \, z \,)^2} \quad .
\end{equation}
If $\alpha + \beta \neq 0$,\vspace{1mm}
\begin{equation}\label{eq:SpecialSolution3.2}
 ds^2 = \left(1 + (\, \alpha + \beta \, ) \, z\, \right)
 ^{\, \frac{2}{3}\left( \frac{3 \alpha - \beta}{\alpha + \beta} \right)} \, dt^2
 - \, \left(1 + (\, \alpha + \beta \, ) \, z\, \right)
 ^{\, \frac{4}{3}\left( \frac{\beta}{\alpha + \beta} \right)} \, \left(dx^2 + dy^2 \right) 
 - \, dz^2 \quad , 
\end{equation}
\vspace{1mm} 
and if $\alpha + \beta = 0$, 
\vspace{1mm}
\begin{equation}\label{eq:SpecialSolution3.3}
 ds^2 = e^{\, \frac{8}{3} \, \alpha \, z} \, dt^2
 - \, e^{\, - \, \frac{4}{3} \, \alpha \, z} \, \left(dx^2 + dy^2 \right) 
 - \, dz^2 \quad .
\end{equation}
\vspace{1mm}
In the special case $p = 0$ we obtain, for $\beta = 0$, the Minkowski metric  described 
by an observer with a uniform acceleration $\alpha$ (\cite{GrHer}\,) 
or, for $\alpha = 0$, the Taub-Levi-Civita vacuum solution (\cite{AmGr}\,) . 
\end{enumerate}

\end{thm}

\proof
If we take the metric (\ref{eq:PlaneMetric}) in coordinates such that 
$du = \sqrt{-g_{33}}\,dz$ and define%
\footnote{This is a further simplification for the form of the metric 
appearing in \cite{DolgKhri}. }
\begin{equation}
 \xi(u) = \frac{3}{4} \, \frac{d}{du}\, \ln \,|\, g_{11}\,|  
 \qquad \text{and} \qquad
 \psi(u) =  \frac{d}{du}\, \left( \, \frac{1}{2} \, \ln \, g_{00} 
         +  \frac{1}{4} \, \ln \,|\, g_{11}\,| \, \right)  \quad ,
\end{equation}
such that
\begin{equation}\label{eq:SolutionMetricDogov}
 ds^2 = e^{{2 \over 3} \int du \,( 3 \psi(u) - \xi(u))} \, dt^2
 - e^{{4 \over 3} \int \,du \, \xi(u)} \, \left(dx^2 + dy^2 \right) - \, du^2 \quad , 
\end{equation}
then Einstein Equations turn into 
\begin{eqnarray}\label{Eq:EinsteinDolgov1}
G^t_t = T^t_t  & : & 
- \frac{4}{3} \left( \xi' + \xi^2 \right) = \rho  \quad \\
\label{Eq:EinsteinDolgov2}
G^t_t - 4 G^x_x = T^t_t - 4 T^x_x  & : & 
4 \left( \psi' + \psi^2 \right) = \rho + 4 p - 4 \delta p \quad \\
\label{Eq:EinsteinDolgov3}
G^u_u = T^u_u  & : & 
- \frac{4}{3} \, \xi \, \psi = - p \quad .
\end{eqnarray}
If the derivative of the potential is zero everywhere then $p=0$. Thus, 
from eq.(\ref{Eq:EinsteinDolgov3}), we conclude that $\xi = 0$ 
or $\psi =0$. 

If  $p=0$ and $\xi =0$, then $\rho = 0$ as eq.(\ref{Eq:EinsteinDolgov1}) demands. 
Defining the coordinate function $z=\psi + z_1$, for an arbitrary constant 
$z_1 \neq \sqrt{-\delta p(0)}$, from eq.(\ref{Eq:EinsteinDolgov2}) we obtain
\begin{equation}
dz = - \left( \, \delta p + (z-z_1)^2 \, \right) \, du \quad ,
\end{equation}
implying formula (\ref{eq:SpecialSolution1}). This proves (i).

If  $p=0$ and $\psi =0$, then $\rho = 4 \, \delta p$ follows from 
eq.(\ref{Eq:EinsteinDolgov2}). As before, defining the coordinate function 
$z=\xi + z_1$ and using eq.(\ref{Eq:EinsteinDolgov1}) we get eq.(\ref{eq:SpecialSolution2}),
hence proving (ii).

If the derivative of the potential is nowhere defined then $\rho  = p$ and 
$\delta p = 2\, p$. From eq.(\ref{Eq:EinsteinDolgov1})-(\ref{Eq:EinsteinDolgov3}), 
we obtain the following system of ODE's:
\begin{equation}
 \xi' + \xi^2 + \psi \, \xi = 0  \qquad 
 \psi' + \psi^2  + \psi \, \xi = 0  \quad . 
\end{equation}
Its general solution is, after we put $z=u$,
\begin{equation}
 \psi = \frac{\alpha }{1 + (\alpha + \beta) \, z}  \qquad 
 \xi = \frac{\beta}{1 + (\alpha + \beta) \, z}  \quad . 
\end{equation}
Returning this expressions in the previous formulas, we prove (iii). 
\qed
%
%

%
%
%
\section{Applications to simple fluids}

To illustrate the theorems above, assume we have equations of state in the form
\begin{equation}\label{eq:LinearEquationState}
 p = (\gamma(\rho)-1)\, \rho   \qquad \text{and} 
 \qquad \delta p = \epsilon(\rho)\, \rho  \quad .
\end{equation}
In this case, the configuration of the system is completely determined solving
the simple first order ODE of hydrostatic equilibrium for $\rho(z)$:
\begin{equation}\label{eq:SelfGravityConservation2}
\left(   \frac{d\gamma}{d \rho} \, \rho + \gamma -1 \right) \, \frac{d\rho}{dz}= - \left( \, 
\frac{ (z-z_1) \,\gamma  +  \,(z_2-z_1)\,(4 \epsilon - \gamma)}{(z-z_1) \,(2-\gamma)  
+  \,(z_2-z_1)\,( 7 \gamma - 4 \epsilon - 6)}  \,\right)\frac{(\gamma-1)\,\rho}{z - z_1} \quad .
\end{equation}
Even though this ODE is not exactly solvable, any approximation method for its solution
reflects instantaneously in the metric components, as demanded  by equations (\ref{eq:GeneralMetric}),
(\ref{eq:DefinitionPotential}) and (\ref{eq:LinearEquationState}). 
This is a great simplification in the analysis of such systems.

 Specializing to constant $\gamma$ and $\epsilon$, such that $\gamma \neq 1$, $\gamma \neq 2$ and 
$7\,\gamma \neq  4\, \epsilon + 6$, eq.(\ref{eq:SelfGravityConservation2}) is easily integrated as
\begin{equation}\label{eq:EnergyCostantCoeficients1}
 \frac{\rho}{\rho_0} \, =  \left( \, 1 - \frac{z}{z_1} \,\right)^{ \,\frac{\gamma - 4\, \epsilon}{7\,\gamma - 4\, \epsilon -6}}
 \, \left( \, 1 -  \frac{z}{\alpha \, z_1} \,\right)^{\,\frac{-6\, \gamma^2 + 4\, \gamma + 8\, \epsilon}{(2-\gamma) (7\,\gamma - 4\, \epsilon -6)}}\quad ,
\end{equation}
where $\rho_0 = \rho (0)$ and  
\begin{equation}\label{eq:EnergyCostantCoeficients2}
\alpha = 1- \left(\, 1-\frac{z_2}{z_1}  \,\right) \,
\left(\, \frac{7\,\gamma - 4\, \epsilon -6}{\gamma - 2} \,\right) \, \quad .  
\end{equation}
Applying it in eq.(\ref{eq:GeneralMetric}), we arrive to the expression for the metric 
\begin{equation}\label{eq:MetricAffine1}
 \begin{array}{c}
 ds^2 =  \, \left( 1 - \frac{z}{z_1} \right)^{\,\frac{2\, (1-\gamma)}{7\,\gamma - 4\, \epsilon -6}}
 \, \left( 1 -  \frac{z}{\alpha \, z_1} \right)^{-\,\frac{2\, (1-\gamma )(6\, \gamma - 4 \, \epsilon-4)}{(2-\gamma) (7\,\gamma - 4\, \epsilon -6)}}
 \,  \, dt^2 \\[6mm]
 -
  \, \left( 1 - \frac{z}{z_1} \right)^{-\frac{4(1-\gamma)}{7\,\gamma - 4\, \epsilon -6}}  
 \, \left( 1 -  \frac{z}{\alpha \, z_1} \right)^{\frac{4(1-\gamma)}{7\,\gamma - 4\, \epsilon -6}}
\, (dx^2 + dy^2) \\[6mm]
-
 \, \frac{4\, (1-\gamma)\,(z_2-z_1)}{\alpha^2 \, (2-\gamma)^2 \, \rho_0 \, z_1^3} 
 \, \left( 1 - \frac{z}{z_1} \right)^{\frac{- 8\, \gamma + 8\, \epsilon + 6}{7\,\gamma - 4\, \epsilon -6}}  
 \, \left( 1 -  \frac{z}{\alpha \, z_1} \right)^{\frac{4(5\, \gamma^2 - 2 \,\gamma\, \epsilon - 11\, \gamma + 2\, \epsilon + 6) }{(2-\gamma) (7\,\gamma - 4\, \epsilon -6)}}
 \, dz^2
 \end{array} 
\end{equation}
with $(1-\gamma)\, \rho_0 \, z_1 > 0$.

Equation (\ref{eq:SelfGravityConservation2}) is also readly integrable if 
$\epsilon \neq \gamma = 2$ or $\gamma \neq 2$ and $7\,\gamma = 4\, \epsilon + 6$,
giving similar results. 
For $\gamma = 1$ or $\epsilon = \gamma = 2$, we must apply theorem 
(\ref{thm:SpecialSolutions}).
Therefore, the whole range of possibilities for constant $\gamma$ and $\epsilon$ are 
easily covered. 

In the literature, the perfect fluid $\epsilon = 0$ is most studied
(\cite{Taub} ,\cite{TaTau}, \cite{TWS}, \cite{BK}, \cite{Col}, \cite{Sar}, \cite{SKMHH}-sec. 15.7.1), 
followed by the Einstein-Maxwell
system of a charged infinite plane (~\cite{AmGr}, \cite{McV}~), 
where $\gamma =0$ and $\epsilon = -1$. The reader is invited to analyse in those works
the effort one had to make solving Einstein equations for very specific 
values of $\gamma$ and $\epsilon$, 
whilst in our approach it is no more than a simple application of 
theorem \ref{thm:GeneralSolution}, which gives general formulas just like 
in eq.(\ref{eq:MetricAffine1}). 

 We could have gone even further inserting a cosmological  constant $\Lambda$  from the 
beginning: just take $\rho + \Lambda$ and $p - \Lambda$ instead of $\rho$ and $p$, 
respectively. As an example, the family of vacuum solutions with cosmological constant
(\cite{NoHors,Hors}) is straightforwardly integrated from 
eq.(\ref{eq:DefinitionPotential}) after such substitution is made and the "new"
observables attained the values \linebreak $\rho = p =\delta p =0$.


\section{Concluding remarks}

 In this work we have shown that any static plane-symmetric metric can be given, at least locally,
as a simple functional of the matter content of space-time. In mathematics, it resembles 
the hard problem of characterizing the metric from its curvature tensor, 
which, in general, is impossible (See \cite{Berger}, sec. 4.5, for a short account 
of this problem in Riemannian geometry and the references therein). Contextualizing to general
Relativity, due to the relevance of the generalized Birkhoff theorem ascribing a local form 
for the metric to vacuum solutions with certain symmetries \cite{SKMHH}, 
theorems \ref{thm:GeneralSolution}
and \ref{thm:SpecialSolutions} form a kind of "Birkhoff-type-theorem" with matter:
\emph{Given a certain class of symmetry assumed by the space-time and its matter content,
decide if it is possible to find local coordinates for which the metric can be put as
an explicit functional of the components of the energy-momentum tensor $T_{\mu\nu}$ 
such that both satisfy Einstein equations
under the sole hypothesis of conservation $\nabla_\mu \, T^\mu_\nu=0$, where the later  
corresponds to self-gravitating equations for the energy and momentum of the system. } 

 The results obtained so far purport to be valuable in the foundations of 
General Rela-\\tivity. As a particular interest, grasping
the relationship between the quantum Casimir effect and gravitation 
was the starting point of this work \cite{dLGM}. From this viewpoint, we have just set the 
ground where the problem may lay, 
and grasping it is a long, deep and uncertain journey. Investigations are under way.


\end{document}